\documentclass{article}%
\usepackage{amsmath}%
\usepackage{amsfonts}%
\usepackage{amssymb}%
\usepackage{graphicx}
\usepackage{tikz}
\usepackage{circuitikz}
\usepackage{psfrag}
\usepackage{caption}
\usepackage{subcaption}
\usepackage{xcolor}
\usepackage[normalem]{ulem}
\begin{document}

\vspace{32pt}
\begin{center}
{\textbf{\Large MEASURING THE HANNAY GEOMETRIC PHASE}}
\vspace{40pt}

H.~Fanchiotti$^1$, C.A.~Garc\'\i a Canal$^1$, M.~Mayosky$^2$, A.~Veiga$^3$, V.~Vento$^4$\\
 \vspace{12pt}
\textit{$^1$IFLP/CONICET and Departamento de F{\'\i}sica}\\ \textit{Universidad Nacional de La Plata, C.C.67, 1900, La Plata, Argentina}\\
\vspace{10pt} \textit{$^2$LEICI, Departamento de Electrotecnia Facultad de Ingenier{\'\i}a}\\ \textit{Universidad Nacional de La Plata, La Plata,
Argentina and}\\ \textit{ Comisi\'on de Investigaciones Cient{\'\i}ficas de la Provincia de Buenos Aires-CICpBA, Argentina}\\ \vspace{10pt}
\textit{$^3$LEICI, Departamento de Electrotecnia Facultad de Ingenier{\'\i}a,Universidad Nacional de La Plata, La Plata,
Argentina; CONICET}\\ \vspace{10pt}
\textit{$^4$Departamento de F{\'\i}sica Te\'orica-IFIC. Universidad de Valencia-CSIC.}\\ \textit{E-46100, Burjassot (Valencia), Spain}

\end{center}

\vspace{40pt}

\date{\today}%

\begin{abstract}

An electric network whose dynamics is entirely equivalent to the Foucault pendulum is presented. This circuit allows one to measure in a simulation of the circuit, and eventually in a built circuit in the laboratory, the Hannay geometric phase. This phase corresponds, via the equivalence discussed in the text, exactly to the latitude effect on the rotation of the oscillation plane of the pendulum. The circuit includes a gyrator, that introduces a non-reciprocal character to the network.
\end{abstract}

\vspace{2cm}

The aim of this letter is to present a measurement of the geometric phase in a classical system. This classical geometric phase, called Hannay angle \cite {Hannay} is the analog to the well-known quantum geometric phase discovered by Berry \cite{berry}. We notice that in Ref.\cite{Xu}
(see also \cite{HannayXu}) a particular circuit for obtaining the Hannay angle was investigated based on time-dependent circuital elements and a scheme for measuring this phase was proposed, wishing for an experimental verification to be done. Another proposal was presented in Ref.\cite{Bhatt}. In this presentation we show an experimental electronic setup where the mentioned phase can be measured in a simulation.

 The standard example of a classical geometric phase is the latitude effect of the Foucault pendulum. This pendulum allowed Leon Foucault in 1851 to demonstrate the Earth's rotation. The pendulum has been much studied and was found that the oscillation plane of the pendulum has a rotation frequency that depends on the latitude where it is located. In our opinion the Foucault pendulum merits the representation in terms of an equivalent electronic circuit that we present here because it provides an easy way of measuring the classical geometric phase.

Let us start by remembering the standard equations that describe the Foucault pendulum and that one can find in any textbook. Calling
$x$ and $y$ the projections of the bob of the pendulum in the plane $(x,y)$ perpendicular to the vertical direction, the equations under the usual approximations are
\begin{equation}\label{x}
\ddot{x}+ 2\,\Omega_T\, \sin(\lambda)\,\dot{y}+\omega_0^{2}\,x=0,
\end{equation}
\begin{equation}\label{y}
\ddot{y} -2\,\Omega_T\, \sin(\lambda)\,\dot{x}+\omega_0^{2}\,y=0,
\end{equation}
where $\Omega_T$ is the Earth rotation frequency, $\lambda$ is the latitude of the pendulum location and $\omega_0$ the oscillation frequency of the pendulum.

\begin{figure}
	\centering
	\begin{circuitikz}
		\draw (4,0)  node[gyrator] (G) {} node[above]{G} ;
		\draw (0,-1.05) to [C,l=$C$] (0,1.05);
		\node at (0.4,0.4) {$V_{C_0l}$};
		\draw (2,-1.05) to [L,l=$L$,*-*] (2,1.05) node[above]{$V_1$};
		\draw (6,-1.05) to [L,l=$L$,*-*] (6,1.05) node[above]{$V_2$};
		\draw (8,-1.05) to [C,l=$C$] (8,1.05);
	  \node at (8.4,0.4) {$V_{C_0r}$};
		\draw (0,1.05) to [switch,l=$t{=}0$] (2,1.05);
		\draw (6,1.05) to [switch,l=$t{=}0$] (8,1.05);
	  \draw (G.B1) to (6,1.05);
		\draw (G.B2) to (8,-1.05);
		\draw (G.A1) to (2,1.05);
		\draw (G.A2) to (0,-1.05);
		\draw (2,-1.05) node[ground]{};
		\draw (6,-1.05) node[ground]{};
	\end{circuitikz}
		\caption{Resonant circuits coupled by a Gyrator}
		\label{fig1}
\end{figure}
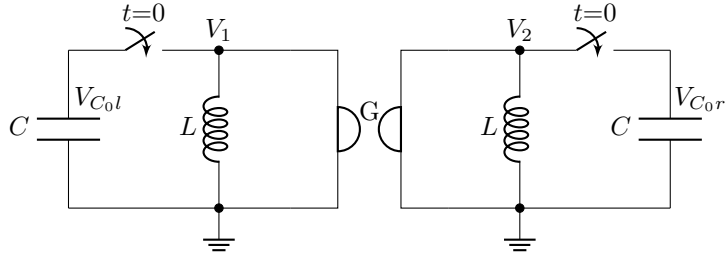
As is well known, an electric network includes a set of elements like capacitors and inductances and connections among them. We are interested in circuits where voltage and current depend only on time and we restrict our analysis to passive networks. The network has ports: pairs of terminals that allow to exchange energy with the surroundings and have a given voltage and current.  We call V the vector corresponding to the port voltage. A network not connected to external energy sources is reciprocal if considering two different terminals, say $\alpha$ and $\beta$, the excitation in $\alpha$ gives rise to a response in $\beta$ that is invariant under the permutation $\alpha \Leftrightarrow \beta$.

In the area of electric networks one can find another classical system that is governed by completely analog equations to those presented above for the  Foucault pendulum. It is the
circuit network including two $LC$ oscillators with the same frequency $\omega_0=1/\sqrt{LC}$, coupled by a so called gyrator as presented in Fig.\ref{fig1}. Due to the fact that any combination of $(L,C)$ elements gives rise to a reciprocal network, in order to have non-reciprocal behavior the introduction of the gyrator, which is a passive element of two ports\cite{gyr} with conductance $G$, is needed.

The analysis of the time evolution of the electric circuit results in a system of linear differential equations with constant coefficients
that are \cite{proc}
\begin{equation}\label{V1}
\ddot{V_1}+\gamma_g\,\dot{V_2}+\omega_0^{2}\,V_1=0,
\end{equation}
\begin{equation}\label{V2}
\ddot{V_2} -\gamma_g\,\dot{V_1}+\omega_0^{2}\,V_2=0,
\end{equation}
where
\[
\gamma_g = \frac{G}{C}\,\,\,;\,\,\,{\omega_0}^2 = \frac{1}{LC}.
\]

Consequently, the differential equations for both systems, Eqs.(\ref{x}) and (\ref{y}) and Eqs. (\ref{V1}) and (\ref{V2}), have exactly the same form. In other words, the role of $x$ and $y$ is taken by $V_1$ and
$V_2$ respectively, by establishing the connection
\[
 2\,\Omega_T\,\sin(\lambda)=\frac{G}{C}\,\,\,\,\,and\,\,\,\,\,\omega_0^2 = \frac{1}{L\,C}.
\]
In the previous expressions, as we have stated before, $G$ represents the conductivity of the gyrator while $C$ and $L$ are the capacity and the inductance of each of the coupled oscillators. In order to obtain the appropriate solutions of the differential equations they have to be implemented with the corresponding initial conditions. We will return to this point below.

 Let us now go to discuss the appearence of geometric phase in our scheme. In this respect, it is worth to remember that, as was mentioned, the motion of the oscillation plane of the
Foucault pendulum acquires a latitude effect in the form of a phase shift
\cite{latitude}. This phase was shown \cite{Fphase} to be no more than an
example of the Hannay phase \cite{Hannay}, the classical analog of the Berry phase. In fact, whenever one performs the analysis of the pendulum motion as  seen from an inertial frame outside the Earth with the Earth's center fixed, it is possible to show \cite{von}
that the geometry of the phase is directly related to the solid angle defined at the center of the
Earth by the parallel of the given latitude.

The analysis of the geometric phase in the case of the Foucault pendulum can be presented as follows.
With the projections $x$ and $y$ of the pendulum movement we form the complex variable $z=x+iy$ that on the basis of
equations (\ref{x}) and (\ref{y})) satisfies
\begin{equation}
\ddot{z}+ 2\,i\,\Omega_T\sin(\lambda)\,\dot{z}+a^{2}\,z=0,
\end{equation}
whose general solution reads
\begin{equation}
z(t)=z_{0}(t)e^{-i\Omega_T\,\sin(\lambda)\,t},
\end{equation}
with
\[
z_{0}(t)=A_{1}e^{ia\,t}+A_{2}e^{-ia\,t},
\]
and $A_{1}$ and $A_{2}$ are constants fixed by the initial conditions.

Then, at $t=0$ one has
\[
z(t)=z_{0}(0)\,\,\,;\,\,\,z_{0}(0)=A_{1}+A_{2},
\]
and we can evaluate the solution for a time $t = T$ the period of rotation of the Earth, namely
\[
T=\frac{2\pi}{\Omega_{T}},
\]
to obtain
\[
z(T)=z_{0}(T)e^{-i2\pi\sin(\lambda)}\,\,\,;\,\,\,z_{0}(T)=A_{1}e^{i2\pi\omega
_{0}/\Omega_{T}}+A_{2}e^{-i2\pi\omega_{0}/\Omega_{T}}.
\]

The pendulum frequency is so large when compared with $\Omega_{T}$, that one can safely consider
\begin{equation}\label{n}
\omega_{0}/\Omega_{T}=n.
\end{equation}

Consequently, the solution has the expression
\begin{equation}
z(T)=z_{0}(0)e^{-i2\pi\sin(\lambda)},
\end{equation}
that clearly shows that the solution $z$ returns to the initial state with an extra phase given by
\begin{equation}
\eta = 2\,\pi\,\sin(\lambda).
\end{equation}

This is the Hannay phase associated to the latitude effect of this classical system. This effect is described by the angle $\Delta \phi$ that the oscillation plane of the pendulum has to travel after a complete rotation of the Earth to complete a full $2 \pi$ rotation in a given latitude which is given by 

\begin{equation} \label{geopen}
\triangle \phi = 2\,\pi - \eta = 2\,\pi(1 - cos(\lambda)).
\end{equation}

 Let us now proceed with the analysis of the circuit of Fig.\ref{fig1}.

\subsection*{Results of the simulation}

As we stated, the electric equivalence comprises two identical resonant circuits connected by a gyrator, as shown in Fig.\ref{fig1}. The behavior of this circuit can be simulated using standard packages such as LT-spice. Although it is not included as a component in the LT-spice standard libraries, an ideal gyrator can be implemented with two voltage dependent current sources in a straightforward way. The plot of $V_1$ versus $V_2$ resembles exactly the time evolution of the $x$ and $y$ projections of the pendulum bob. Initial conditions for simulation are imposed as initial charges in each capacitor. This guarantees that this initial conditions represent the Foucault pendulum when it starts with zero initial velocity.

\begin{figure}
	\centering
	\begin{subfigure}[b]{0.45\textwidth}
         \centering
				\psfrag{V1}{$V_1$}
	      \psfrag{V2}{\raisebox{-5pt}{$V_2$}}
         \includegraphics[width=\textwidth]{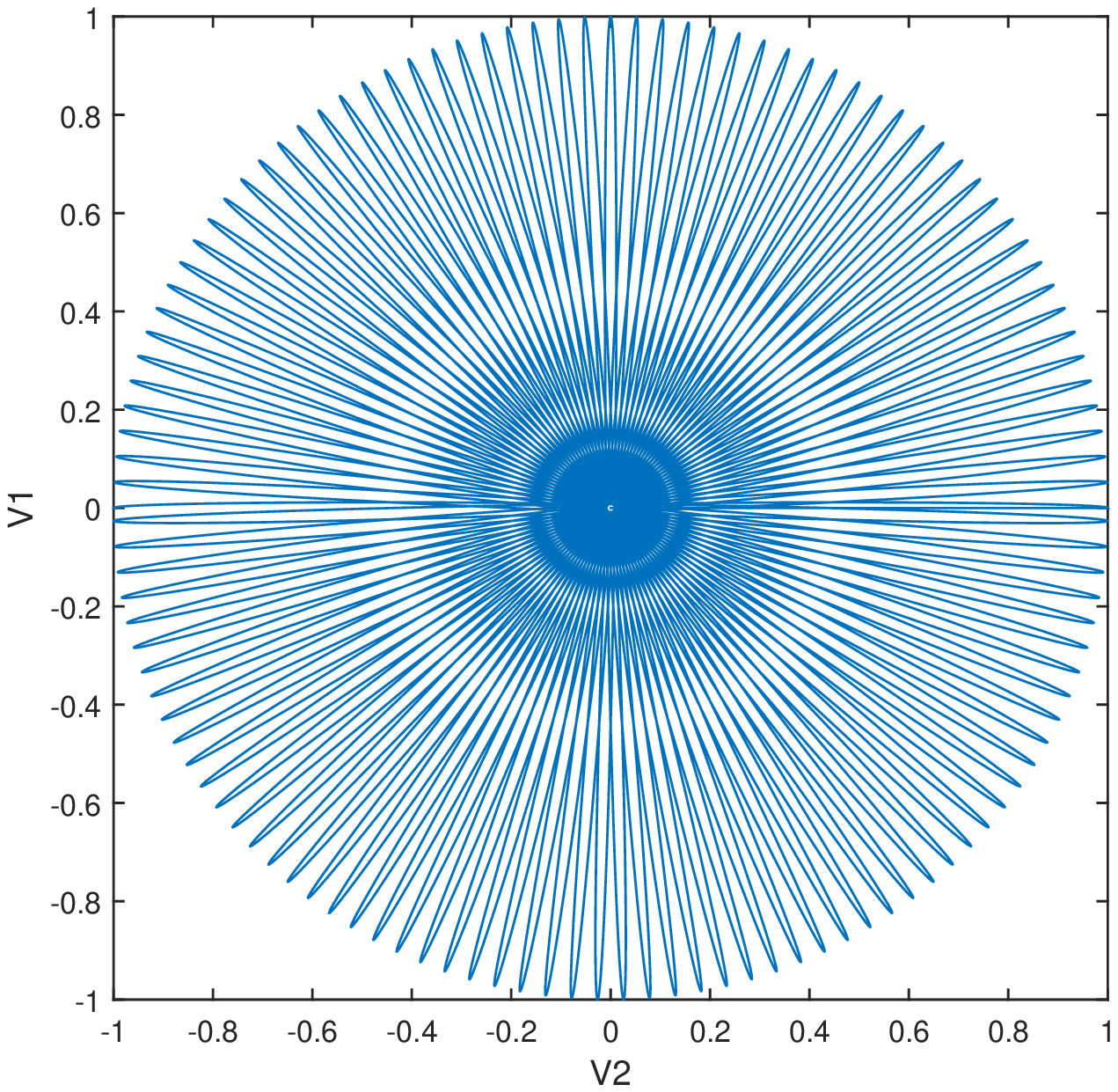}
     \end{subfigure}
\hfill
		\begin{subfigure}[b]{0.45\textwidth}
         \centering
				\psfrag{V1}{$V_1$}
	      \psfrag{V2}{\raisebox{-5pt}{$V_2$}}
         \includegraphics[width=\textwidth]{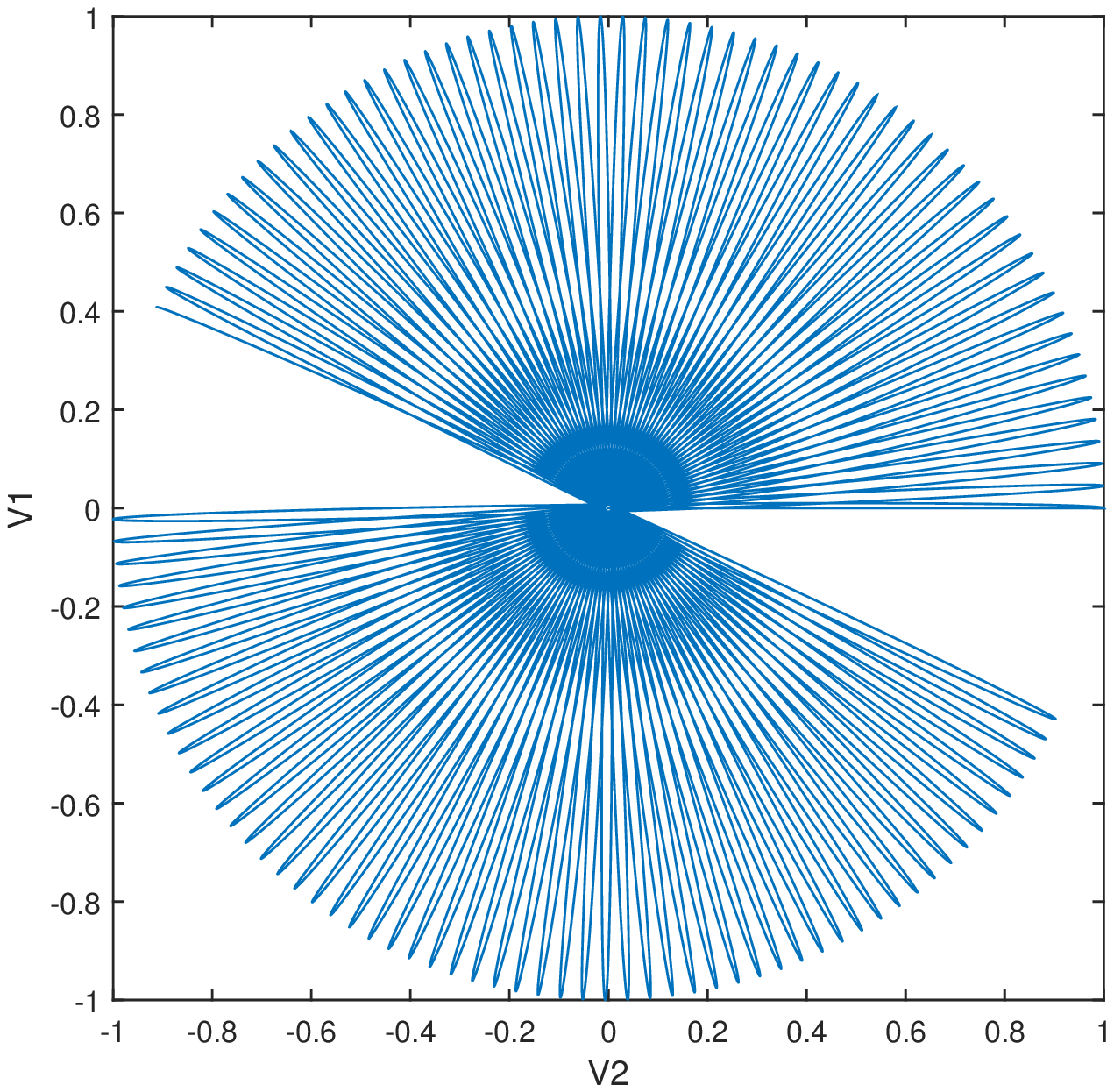}
     \end{subfigure}
	   	\caption{Simulation results. (a) latitude $90^{\circ}$. (b) latitude $60^{\circ}$}
	\label{simus}
\end{figure}

The geometric phase equivalent is now
\begin{equation}
\eta=\frac{\pi}{\Omega_T C}G,
\end{equation}
 so that different latitudes can be simulated just by changing the value of $G$ multiplying it times a value in between $0$ and $1$. In Fig.\ref{simus}, we show the results of the simulations that correspond in the case of the pendulum to latitudes of $90^{\circ}$ (pole) and $60^{\circ}$. Notice that for clarity we decided to show only half of the $2\,\pi$ rotation (half of a day). For these simulations, the circuit values are the following:
\begin{eqnarray}
L &=& 10 mHy \nonumber \\
C &=& 0.1 \mu F   \nonumber \\
G &=& 52.354 \mu S * \sin(latitude) \nonumber
\end{eqnarray}

Initial conditions for both cases are $V_{C_0l}=1V$ and $V_{C_0r}=0V$. Initial currents on inductors are both null. For these circuit values a 12 hour pendulum movement corresponds to 12 milliseconds of simulation time.

It is worth mentioning that in an experimental circuit implementation unavoidable component losses arise. In such a case, the dissipation due to the resistances involved should certainly be taken into account. This implies
the modification of the equations (\ref{V1}) and (\ref{V2}), and both the simulation and the experimental implementation get more involved. For pedagogical purposes, we postpone the treatment of these complexities to a forthcoming publication.

In conclusion, the measurement of the geometric phase of a classical system, the Hannay angle, is easily realized in a simulation of an electric network including a gyrator. This determination has direct correspondence with the so-called latitude effect in the rotation of the oscillation plane of the Foucault pendulum. To summarize,  besides having described a classical system where the geometric phase can be measured, we have presented a complete electric analog of the Foucault pendulum  and have made a simulation that allows to measure geometric angles.

\section*{Acknowledgement}
This work was supported in part  by the MICINN and UE Feder under contract FPA2016-77177-C2-1-P.

\end{document}